\documentclass[reprint, amsmath,amssymb,
 aps]{revtex4-2}

\usepackage{amsmath,amssymb,graphicx}
\usepackage[final]{hyperref}
\usepackage{mathrsfs}
\usepackage{dcolumn}
\usepackage{bm}

\hypersetup{
colorlinks=true, 
linkcolor=blue, 
citecolor=blue, 
filecolor=magenta, 
urlcolor=blue
}

\begin{document}
\title{Low-Light Shadow Imaging using Quantum-Noise Detection with a Camera}
\author{Savannah L. Cuozzo}
\affiliation{Department of Physics, William \&
Mary, Williamsburg,
Virginia 23187, USA}
\author{Pratik J. Barge}
\affiliation{Department of Physics,
Louisiana State University, Baton Rouge, Louisiana 70803, USA}
\author{Nikunjkumar Prajapati}
\affiliation{Department of Physics, William \&
Mary, Williamsburg,
Virginia 23187, USA}
\author{Narayan Bhusal}
\affiliation{Department of Physics,
Louisiana State University, Baton Rouge, Louisiana 70803, USA}
\author{Hwang Lee}
\affiliation{Department of Physics,
Louisiana State University, Baton Rouge, Louisiana 70803, USA}
\author{Lior Cohen}
\affiliation{Department of Physics,
Louisiana State University, Baton Rouge, Louisiana 70803, USA}
\author{Irina Novikova}
\affiliation{Department of Physics, William \&
Mary, Williamsburg,
Virginia 23187, USA}
\author{Eugeniy E. Mikhailov}
\email[eemikh@wm.edu]{}
\affiliation{Department of Physics, William \&
Mary, Williamsburg,
Virginia 23187, USA}

\date{\today}

 \begin{abstract}
We experimentally demonstrate an imaging technique based on quantum noise
modification after
 interaction with an opaque object. 
 By using a homodyne-like detection scheme, we eliminate the detrimental
 effect of the camera's dark noise, making this approach particularly attractive for imaging scenarios that require weak illumination.
Here, we reconstruct the image of an object illuminated with a squeezed
vacuum  using a total of 800
photons, utilizing less than one photon per frame on average. 
\end{abstract}

\maketitle
 
Quantum imaging~\cite{AppsofQuantImagingReview,2photonimagingReview,Berchera_2019,
PhysRevLett.109.123601,imagingwithquantumlight} is capable of outperforming classical alternatives since it's able to utilize
non-classical correlations in probing optical fields.  
Several quantum-enhanced imaging methods have been developed and proved useful for biological imaging~\cite{TAYLOR20161,quantumEnhancedMicroscopy} and imaging in the presence of contaminating classical background illumination~\cite{quantumDistillation,quantum_illumination2020}.
When imaging in the low-photon regime, 
it can be difficult to implement direct intensity detection. This is due to the accuracy of such detection being determined by the photon statistics and by technical noise. Some examples may include laser intensity 
fluctuations or the detector dark noise, and normally requires a long exposure time to 
allow for statistical averaging.

 We experimentally demonstrate an imaging technique
 based on detecting the quantum noise distribution of the quadrature-squeezed vacuum before and after it
 interacts with an opaque object. Our homodyne-like detection scheme allows elimination of the detrimental 
effects of the camera's dark
noise and, potentially, is immune to the classical background illumination while keeping the probing intensity low. This approach is particularly attractive 
for applications requiring weak illumination since the squeezed vacuum inherently has very few
photons illuminating the object. 

Many recent realizations of quantum imaging use two-mode optical fields with correlated
 intensity fluctuations generated either through parametric down
 conversion~\cite{subshotspatialcorPRL,genovesePRL2009,spatial_cor_pdc2010,
multimodetwinGenovese2011} or four-wave mixing in an atomic vapor
~\cite{lettSci08,marinoPRA,ImagingSpatEntanglement,marinoPRA16,
MarinoPhysRevA.100.063828}. When an object is placed in one of the optical beams,
 its shape can be imaged with sub-shot-noise accuracy by subtracting the
 intensity images of the two quantum-correlated beams \cite{ExpSubShotImaging}. However, the average intensity of each beam  limits the acceptable level of the dark noise.
 Compared with typical photon-counting detectors, CCD cameras often present a challenge
 for imaging weak optical fields due to their relatively slow frame rate, making it harder to
 mitigate low-frequency technical noises) and their intrinsic dark noise~\cite{MarinoPhysRevA.100.063828,
MarinoPhysRevA.98.043853}.

\begin{figure}[t]
\centering
\includegraphics[width =1\linewidth]{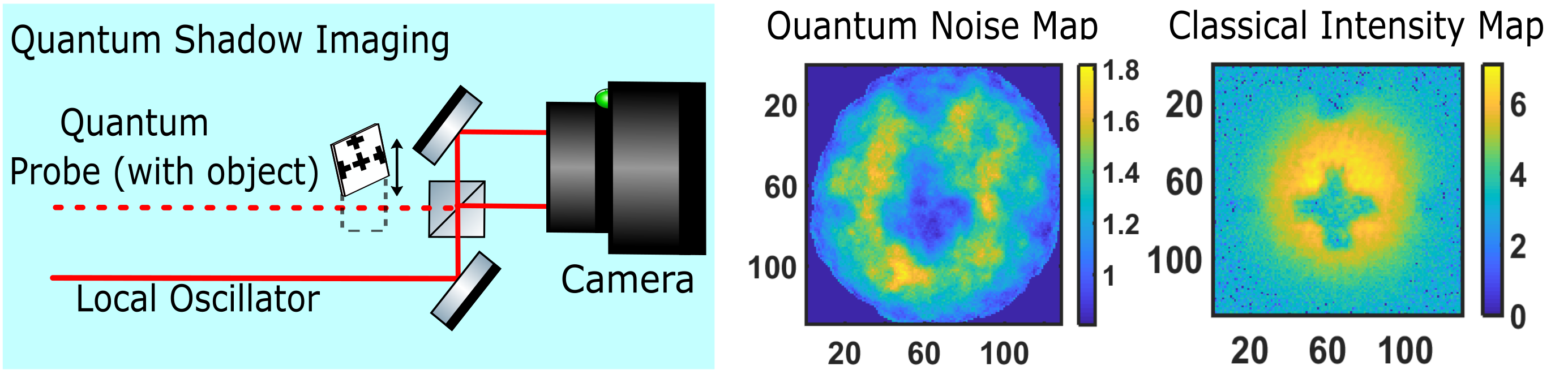}
\caption{
A conceptual representation of the proposed quantum shadow imaging using a "+" as the target.
 The quantum shadow method uses the average quantum fluctuations of the probe and reference
 fields amplified by a local oscillator and therefore not susceptible to the camera's dark noise. The
 quantum shadow probe map is on a linear scale. For comparison, we show classical intensity
 image of the ``+'' target illuminated with a bright beam on a log scale. The ``+'' is about
 475 $\mu$m in width. 
}
\label{fig:concept}
\end{figure}

Our approach is different. Instead of using quantum enhanced  or quantum
correlated intensity measurements, our measurements are based on an analysis of
quantum quadratures variance.
We use a quadrature squeezed vacuum field~\cite{matsko_vacuum_2002,
mikhailov2009jmo,mizhang2016pra,zhangPRA2017}, containing very few photons on
average; when such a field interacts with an opaque object, its quantum
fluctuations in the obstructed zone are replaced with a regular vacuum. To
record the spatial distribution of the resulting noise quadrature without being
affected by the camera dark noise, we mix the quantum probe with a
classical local oscillator field. This amplifies the probe's quantum noise, realizing a
camera-based balanced homodyne detection scheme. Our approach allows us to
image the fields with as low as one photon per frame and yet obtain spatial
details of the object with significantly less acquisition time, making it
attractive to e.g. non-destructive imaging of biological
samples~\cite{quantimaginginacell}. Moreover, in such a method, we can use
an anti-squeezed quadrature --- increasing the tolerance to optical losses.

The concept of the proposed method is illustrated in Fig.~\ref{fig:concept}.
A CCD camera detects the number of photons incident on each pixel, $N$, 
on top of its internal dark noise $N_d$. For a standard intensity
measurement, the boundary between a fully illuminated region (the average
photocounts $\langle N + N_d\rangle$) and a fully blocked region (the
average photocounts $\langle N_d\rangle$) can be distinguished by the
difference between these two photocount values. Moreover, we can estimate
the signal-to-noise of such traditional measurements as 
\begin{equation} \label{eq:tmap_t_snr}
\text{SNR}_t = \frac{\bar{N}}{\sqrt{\bar{N} + 2(\Delta N_d)^2}},
\end{equation}
where $\bar{N}$ is an average photon number detected per pixel (or bin),
 and $\Delta N_d$ is the standard deviation of the dark noise counts.

We propose instead to measure the normalized variance, $V$, of the
quadrature $X_{\theta} = \cos(\theta)X_1+\sin(\theta)X_2$, where $X_1 = a
+ a^{\dagger}$, $X_2 = i(a^{\dagger} - a)$, and $a^{\dagger}(a)$ is the
creation(annihilation) operator for the photon state. In this case a similar
boundary between the light and darkness can be detected via the deviation
of the noise variance for the region illuminated by a quantum probe from
unity --- the noise variance of the coherent vacuum (shot noise). This method 
does not work for a coherent illuminating state because the quadrature variance
is unchanged by the loss. 

For example, if an
experiment uses a squeezed vacuum with the squeezing parameter $r$, the
expected variance value for the squeezed and anti-squeezed quadratures are
$V=e^{\mp2r}$, respectively. We can also estimate the noise of such
measurements by calculating the variance of the corresponding variance
values for such a squeezed vacuum field, yielding the following theoretical
signal-to-noise ratio:
\begin{equation} \label{eq:tmap_q_snr}
\text{SNR}_q = \frac{V-1}{\sqrt{2+2V^2}}
\end{equation}

Note, that for this calculation we can neglect the camera dark noise thanks
to the homodyne detection.
As a result, we can compare the performance of the two approaches as a
ratio of the two signal to noise values for an anti-squeezed vacuum field, and
a coherent beam with similar average number of photons $\bar{N} =
\sinh^2(r) \ll 1$:

\begin{eqnarray}
\label{eq:SNRdn}
\frac{\text{SNR}_q}{\text{SNR}_t} &=& \frac{e^{2r}-1}{\sqrt{2+2e^{4r}
}}\frac{\sqrt{\sinh^2(r)+2(\Delta N_d)^2}}{\sinh^2(r)}\\
\nonumber
& \simeq& \sqrt{1+\frac{2(\Delta N_d)^2}{\bar{N}}}
\end{eqnarray}

It is easy to see that in the limit of the small photon number $\bar N \ll 1$,
the two methods
perform equally well in the case of vanishing dark noise; however, if the dark
noise becomes comparable with the average photon number, the advantage
of the quantum noise-based measurement becomes more obvious.

With our method, we can produce a quality transmission map from
the quantum noise measurements and avoid the detrimental
 effect of the dark noise. We connect the measured
field variance, $V(\vec{x})$ to the object transmission, $T(\vec{x})$ (see Ref.
\cite{Bennink_Boyd_PhysRevA} and
supplementary materials
for detailed derivations):
\begin{equation}
\label{eq:VvsT}
V(\vec{x}) = 1 + (e^{\pm 2r}-1)|\mathcal{O}(\vec{x})|^2 \times T(\vec{x}).
\end{equation}
where $\mathcal{O}(\vec{x}) = \int_A u_{\mathrm{LO}}u_{SqV}^{*}\mathrm{d}A$
is the overlap between the spatial modes of the local oscillator,
$u_{\mathrm{LO}}$, and the squeezed vacuum mode, $u_{SqV}$, and $A$ is the pixel at location $\vec{x}$.
For the reference beam, where the object is removed, we assume $T=1$
everywhere. For the mode-matched local oscillator and quantum probe,
we arrive at the following expression of the transmission map using
measured quadrature noise variance $V_p$ and $V_r$ in the probe and
reference beams, respectively:
\begin{equation}
\label{eq:tmap}
T_q(\vec{x}) =\frac{V_p(\vec{x})-1}{V_r(\vec{x})-1}.
\end{equation}
Note that our method of
transmission calculation is agnostic to the choice of the squeezed or
anti-squeezed quadrature. In this experiment, we work with anti-squeezed
quadrature, since it is more robust to the optical losses.

\begin{figure}
\centering
\includegraphics[width = 1\linewidth]{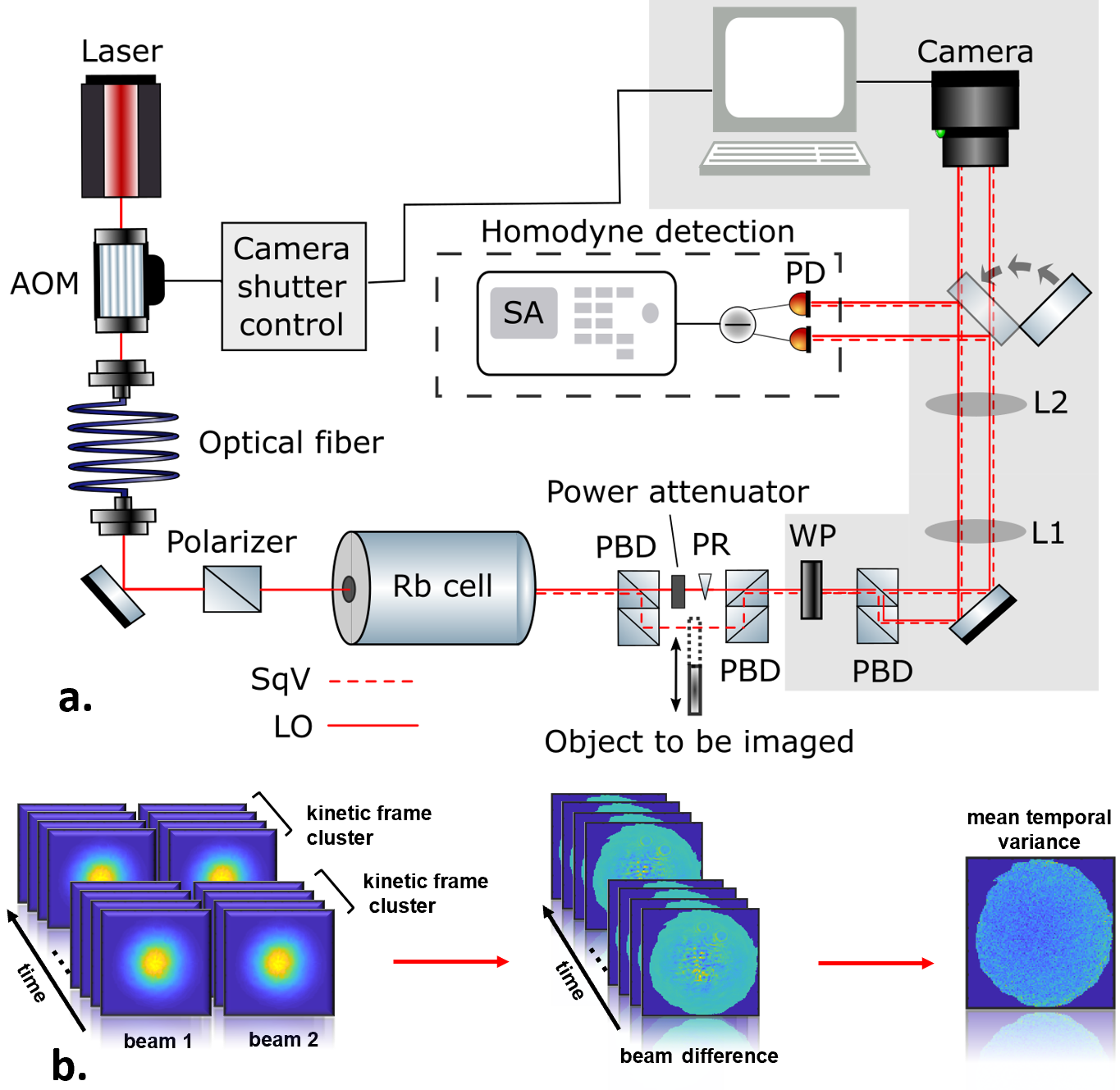}
\caption{a) Experimental setup with two different detection
schemes: traditional homodyne and camera. SqV denotes the squeezed vacuum,
LO denotes the local oscillator, PR is phase retarder,
AOM is an acousto-optical modulator, and
PBD is a polarizing beam displacer. Objects may be placed in the path of
the squeezed vacuum where lenses L1 (300 mm) and L2 (250 mm) map the object image onto the
camera. PDs are photodiodes, SA is a spectrum analyzer, the camera is
connected to a computer. b) Visual illustration of our data analysis. }
\label{fig:expsetup}
\end{figure}

The schematics of the experimental realization of the proposed method is shown in
Fig.~\ref{fig:expsetup}a.
While the specific method of the squeezed vacuum generation is not
important, in the presented experiments
we use a squeezer based on the polarization
self-rotation in a $^{87}$Rb vapor
cell~\cite{matsko_vacuum_2002,mikhailov2009jmo}, details of which are
reported at Ref.~\cite{mizhang2016pra,zhangPRA2017}. The principle difference from
 the previous experimental arrangement is the pulsed
squeezer operation. To avoid camera over-exposure, the pump field is turned
on for only $1~\mu$s during the $544~\mu$s duty cycle using an acousto-optical modulator (AOM).
Right after the squeezer, we detect 1.5~dB of squeezing
and 10~dB of anti-squeezing and these parameters are not
affected by the pulsed operation.
Due to optical losses, after the imaging system we detect (with homodyning
photodiodes) only 0.5~dB squeezing and 7.5~dB anti-squeezing.

After the squeezer, the pump and squeezed vacuum (SqV) fields are physically separated using
a polarizing beam displacer (PBD).
SqV alone passes through the object and then recombines with an attenuated pump field,
which now serves as a local oscillator (LO) in the balanced homodyne
scheme for imaging.
We image the object onto the camera using a 4-f system of lenses (see L1 and L2 in
Fig.~\ref{fig:expsetup}a.

We obtain quantum-limited statistics from images of the two beams
using a Princeton Pixis 1024 camera that has 13$\mu m\times$13$\mu m$ pixels, an average standard 
deviation of dark noise counts of 10 per pixel and high
quantum efficiency (above 95\%), cooled to -70$^{\circ}$C. We illuminate our object with an average of $6\times 10^{-5}$ photons per pixel per frame, so we are in the regime where the dark noise is significantly larger than the photon number. Hence, our quantum method has an advantage according to Eq.~\ref{eq:SNRdn}. 
This camera can only rapidly capture four frames before having to pause for half a second for data
transfer. Thus, we collect four frames, separated by $544 \mu$s (synchronized with the pulsed laser) 
that form "kinetic clusters".
To extract the information about the quantum noise variance, we subtract the intensities of the two 
beams after the final beam splitter (labeled ``beam 1'' and ``beam 2'' in \ref{fig:expsetup}b) to create an amplified noise map
- a 2D analog of the differential photo-currents in a
traditional homodyne detection scheme. 
Next, we calculate the image of the experimental quantum variance $V_{\mathrm{exp}}^{(\mathrm{R})}(x,y)$ normalized to the shot
noise and temporally-average over a given kinetic cluster:

\begin{equation}
\label{eq:VarExp}
V_{\mathrm{exp}}^{(\mathrm{R})}(\vec{x}) = \frac{\left\langle \left( N^{(\mathrm{R})}_{\mathrm{1}}(\vec{x})-N^{(\mathrm{R})}_{\mathrm{2}}(\vec{x})\right)^2\right\rangle}{\left\langle N^{(\mathrm{R})}_{\mathrm{1}}(\vec{x})+N^{(\mathrm{R})}_{\mathrm{2}}(\vec{x}) \right\rangle}
\end{equation}
where the average
is taken within the four frames of each kinetic cluster. Finally, we
average the variance maps over all the kinetic clusters for a given set of experimental
parameters to produce an average normalized quantum noise map of our squeezed
vacuum.
\begin{figure}[t]
\centering
\includegraphics[width =1\linewidth]{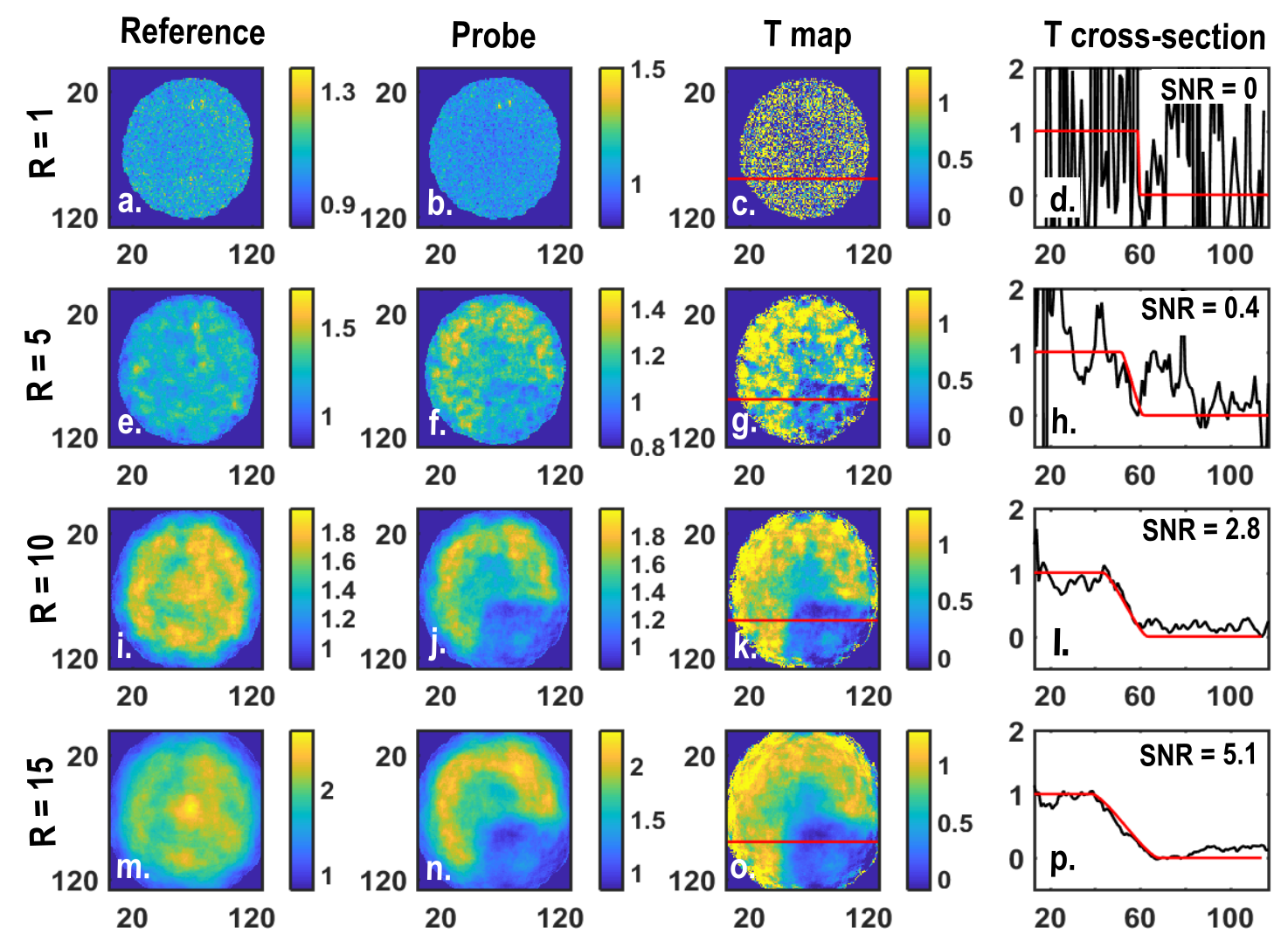}
\caption{
The first column (a, e, i, m) shows the noise maps (defined with
Eq.~\ref{eq:VvsT})
of
the squeezed vacuum with no objects in the vacuum port. The second column
(b, f, j, n) shows the noise maps for the squeezed
vacuum with the mask in the vacuum port. The third column
(c, g, k, o) shows the transmission map, T, as defined in
Eq.~\ref{eq:tmap}. In the final column, the black line shows transmissions
through the mask in the T maps where the red line contrasts the
noiseless classical intensity cross-section through the same region for the same
binning. The first row (a, b, c, d) is binned with a radius of R = 1, the second row (e, f, g, h)
 is binned with a radius of R = 5, the third row (i, j, k, l) is binned with a radius of R = 10, and the last row (m, n,
o, p) is binned with a radius of R = 15. The reference and probe maps are on a linear noise scale. 
The knife edge inserted into the beam is about 0.65 mm $\times$ 0.65 mm.
}
\label{fig:maps}
\end{figure}

To experimentally demonstrate the capabilities of quantum shadow imaging with
the squeezed vacuum, we chose a completely opaque rectangle as our mask to block
approximately one quadrant of the probe beam as our test object to be inserted only
in the squeezed vacuum channel (see
Fig.~\ref{fig:expsetup} a).
For most measurements, we also need to increase the 
effective detection area to  improves the overlap parameter (see Eq.~\ref{eq:VvsT}) with the characteristic quantum-mode size of the squeezed vacuum beam. To do that, for each point, 
$\vec{x}=(x,y)$, we sum all the counts in the radius ${\mathrm{R}}$ (in units of pixels) around 
it to calculate the total photon counts $N^{(\mathrm R)}_{\mathrm{1,2}}(\vec{x})$ -- a process
 commonly refer to as ``binning''. 
 The situation in which the detection area is much smaller than the mode size of the squeezed vacuum is equivalent to a large optical loss, and thus reduces any non-classical noise down to shot noise. 
 Note, this summation is very different
 from having large pixel, since the quantum uncertainty of detection within
 one pixel causes the integration over the field’s amplitude across the
 pixel. While summing over pixels integrates over intensity (or photon
 number), this is correct only when the avalanche process is unlocalized.
 When the avalanche is localized even within one pixel the integration is
 over photon number~\cite{VLPC}.
 
Fig.~\ref{fig:maps} shows the examples of measured variance maps for both
reference and probe beams for different binnings.  
Fig.~\ref{fig:maps} (column four)
shows a cross-section of the experimental quantum shadow transmission map at the location of the red line and
compares it with the calculated transmission map of
an ideal noiseless beam sampled with the same binning of radius R.
When the
radius (R) of the bin is small (top) it is impossible to see
the quantum shadow, since the detected quantum statistics is indistinguishable
from a shot-noise limited beam~\cite{cohenPRL20}. However, as we increase the radius of the bin
(top to bottom rows), the difference in quantum statistics between the blocked and
open regions of the mask becomes more and more pronounced, creating a resolvable
``quantum shadow''. Such improvement, however, comes with the price of
somewhat reduced ``sharpness'' of the image features.
This is because the spatial resolution of the quantum
noise maps is inversely proportional to the size of the bin,
while the contrast of the edge is
proportional to the bin. 

The spatial resolution is also tied to the size of the squeezed mode~\cite{marinoPRA16}, as seen in Eq.
(\ref{eq:VvsT}), since the
size of the bin needs to correspond to the size of the mode for
the best contrast. Thus, in general, a multimode squeezed field with a small mode size
is more attractive for imaging applications, compared to a single-mode optical field.
Some information about the mode decomposition of our squeezed vacuum field may
be gleaned from the first column of images in Fig.~\ref{fig:maps}. If our reference
beam was in the single-mode matching the LO, we would expect it to have a normalized variance
proportional to the overlap parameter of a fundamental Gaussian spatial
mode with itself according to Eq.~\ref{eq:VvsT}.
However, a clear ring-like structure emerges as we
increase the binning radius, suggesting the presence of weaker higher-order modes.
Nevertheless, our close to single-mode squeezer demonstrates quite
good visibility of the image.

\begin{figure}[t]
\centering
\includegraphics[width =1\linewidth]{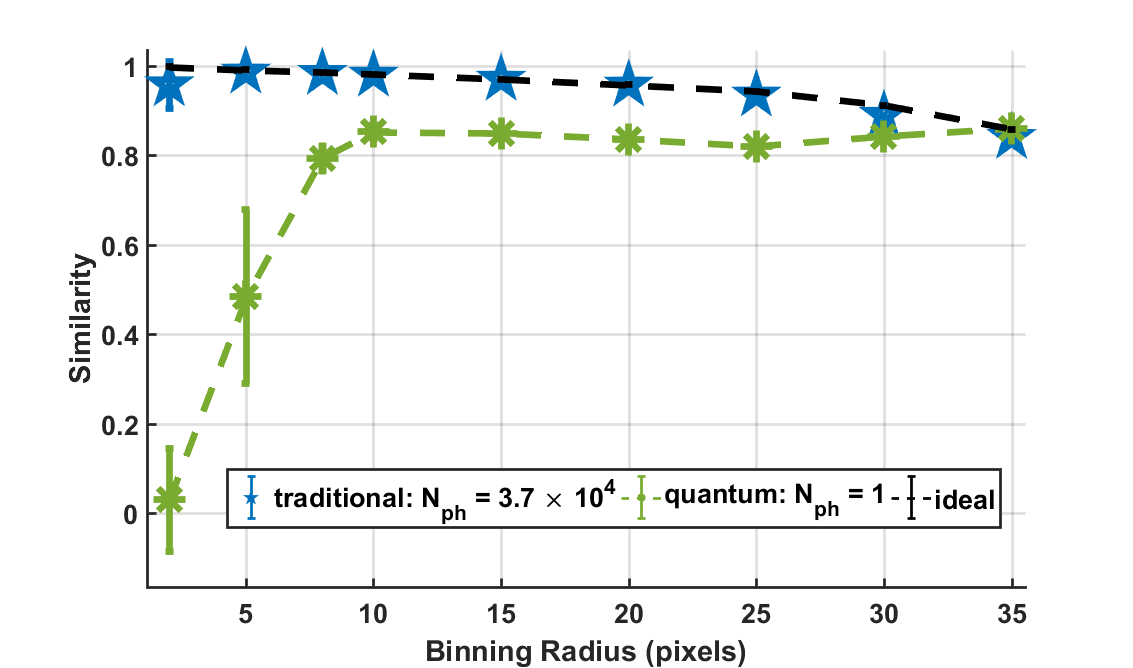}
\caption{
The similarity of our imaging method as a function of detection
radius. Each data set corresponds to a different total photon number
($N_{\mathrm{ph}}$) used to illuminate the image. The black dash-dotted line
shows the best similarity possible with our method assuming a perfectly
noiseless image. We estimate $N_{\mathrm{ph}}
^{\mathrm{quant}} = N_{\mathrm{sq}}\times t_{\mathrm{expo}}/
t_{\mathrm{coherence}}$, where $N_{\mathrm{sq}}=1$ is the number of
photons in a squeezed mode with 7.5 dBs of anti-squeezing (assuming a single mode),
$t_{\mathrm{expo}} = 2\times 10^{-6} s$ is the exposure time of a
frame and $t_{\mathrm{coherence}} = 2.5\times 10^{-6} s$ is the coherence
time of the squeezing.
The similarity was calculated over a 80-pixel span
centered around the edge of the mask.
}
\label{fig:similaritycomb}
\end{figure}

To quantify the quality of our quantum noise images, we calculate the
similarity defined as
\begin{equation}
\label{eq:similarity}
S = \frac{\sum{T_{\mathrm{exp}}T_{\mathrm{o}}}}{\sqrt{\sum{T_{\mathrm{exp}}^2}\sum{T_{\mathrm{o}}^2}}},
\end{equation}
where $T_{\mathrm{exp}}$ is the experimentally measured transmission,
$T_{\mathrm{o}}$ is the true object transmission, and the sum is taken for
pixels along a path across the image (we use horizontal straight line
shown in red in Fig.~\ref{fig:maps}d). This metric allows us to
quantify how well our noise analysis reconstructs the image of an object.
We see that the quantum noise images quickly approach the ideal
similarity (see Fig.~\ref{fig:similaritycomb}) and reflects overall the
mask shape well for significantly lower photon numbers (we estimate that
we have about 1 photon per frame in the squeezed vacuum field).
This is because we can boost our quantum noise above the
dark noise using a homodyne-like detection scheme and our squeezed photons
have correlations that allow us to reconstruct the image from the noise
using less object illuminating resources (photons). It is difficult to
compare the noise shadow imaging method to other quantum imaging methods, because they focus on
enhancing pre-existing techniques and comparing SNRs, but {\em  our methods has
no direct classical counterpart capable to operate at such low 
illumination and high dark count noise levels}.


In conclusion, we can image an opaque object by illuminating it
with a squeezed vacuum. Our scheme can use anti-squeezed quadrature which
makes the whole method more robust against optical and detection losses.
We can reconstruct the object by analyzing
the quantum noise statistics that change spatially depending on the mode
structure of the squeezed vacuum and the object. This has application to
any imaging scenario where a high photon number could damage the object, such
as biological imaging. Also, the overall scheme is quite simple and
uses $6\times 10^{-5}$  photons per pixel per frame. We used only 1600
photons in total to reconstruct the object --- far less than other low photon
methods \cite{imagingwithasmallphotonMorris}. We also note that this method
has the potential to be generalized to other quantum states, e.g. a thermal
state since it only depends on the state's deviation from the shot noise.
Since our method is based on analysis of the quantum state variance, it is
potentially immune to the parasitic illumination by the classical light
sources for which the quadrature variance is independent of transmission.

\begin{acknowledgments}
We would like to thank the late Jonathan Dowling for his work throughout this project. We also thank Morgan Mitchell for helpful discussions and comments. This research was supported by Grant No. AFOSR FA9550-19-1-0066
\end{acknowledgments}

\appendix

\section{Quantum homodyning signal at a pixel}
\label{noiseVsT}
The following analysis is based on the assumption that both squeezed state and local oscillator are initially within a single spatial mode.
Initial state, $|\Psi_{int}\rangle$, is generated from the vacuum by the squeezing operator ($\hat{S}$) in mode 1 and the displacement operator ($\hat{D}$) in mode 2. Object to be imaged is placed in mode 1, and is illuminated solely by squeezed vacuum. Corresponding state $|\Psi_{obj}\rangle$ is generated by the action of the object operator ($\hat{T}$) on the initial state. $|\Psi_{bs}\rangle$ is the state obtained after mixing squeezed light and the local oscillator on a 50:50 beam splitter ($\hat{B}$). To calculate variance in the photon number difference at each pixel, we shift from the Hermite-Gaussian mode basis to the pixel basis. This basis transformation is implemented by the operators $\hat{U}_{1,2}(\vec{x})$ to give final state, $|\Psi\rangle$.
\begin{eqnarray}
&&|\Psi_{int}\rangle = \hat{D}_2(\alpha)\hat{S}_{1}(\xi) |0,0\rangle\\
&&|\Psi_{obj}\rangle = \hat{T}_1(\vec{x})|\Psi_{int}\rangle \\
&&|\Psi_{bs}\rangle = \hat{B}_{12}|\Psi_{obj}\rangle \\
&&|\Psi\rangle =\hat{U}_2(\vec{x})\hat{U}_1(\vec{x})|\Psi_{bs}\rangle,
\end{eqnarray}
Calculation of the theoretical variance $\mathscr{V}_{th}(\vec{x})$ gives,
\begin{eqnarray}
&&\mathscr{V}_{th}(\vec{x}) = \left \langle \Psi\left| \left(\hat{N}_1(\vec{x}) -\hat{N}_2(\vec{x})\right )^2 \right|\Psi \right \rangle =\\
&&|\alpha|^2\:|U_{2}(\vec{x})|^2+|\alpha|^2\:(e^{2r}-1) \:|U_{2}(\vec{x})|^2\: |U_{1}(\vec{x})|^2,
\label{eq:lsuV} 
\end{eqnarray}
and normalizing it with the intensity of the local oscillator (as outlined in Eq.~(6) of the main text) gives normalized variance $V(\vec{x})$ as,
\begin{equation}
V_{th}(\vec{x}) = \frac{\mathscr{V}_{th}(\vec{x})}{|\alpha|^2\:|U_{2}(\vec{x})|^2} = 1+ (e^{2r}-1)\:|U_{1}(\vec{x})|^2,
\label{eq:lsuVn}
\end{equation}
where $\hat{N}_{1,2}$ is photon number operator and $r$ is the squeezing parameter.
\section{Quantum variance of a composite detector}

Binning procedure entails summing the values of all of
the neighbouring pixels inside the detection area (A) of binning radius, $R$.
Normalized variance is calculated after applying binning procedure to beam
difference and beam sum matrices. Under the condition that the pixel size
is smaller than $R$ and $\vec{x}'\in \chi$ with $\chi = \left \{\vec{x}': |\vec{x}-\vec{x}'|\leq R \right \}$, it can be written as
\begin{widetext}
\begin{eqnarray}
\mathscr{V}^{(R)}_{th}(\vec{x}) = \left\langle \Psi\left| \left(\sum_{\vec{x}\,'}\hat{N}_1(\vec{x}\,') -\hat{N}_2(\vec{x}\,')\right)^2 \right|\Psi\right \rangle &&=\sum_{\vec{x}\,'}\sum_{\vec{x}''\neq \vec{x}\,' }\left\langle \Psi\left | \left(\hat{N}_1(\vec{x}\,') -\hat{N}_2(\vec{x}\,')\right) \left(\hat{N}_2(\vec{x}\,'') -\hat{N}_2(\vec{x}\,'')\right) \right|\Psi\right\rangle\nonumber\\
&& \qquad \qquad + \sum_{\vec{x}\,'}\left \langle \Psi\left| (\hat{N}_1(\vec{x}\,') -\hat{N}_2(\vec{x}\,'))^2 \right|\Psi \right \rangle.
\end{eqnarray}
\end{widetext}

Substituting the result of Eq.(\ref{eq:lsuV}) for the second term, followed by expansion of the first term yields
\begin{widetext}
\begin{eqnarray}
\mathscr{V}^{(R)}_{th}(\vec{x}) &&= \sum_{\vec{x}\,'}|\alpha|^2\:|U_{2}(\vec{x}\,')|^2\left ( 1+(e^{2r}-1)|U_{1}(\vec{x}\,')|^2 \right )+ \left \{\left | \sum_{\vec{x}\,'}^{} U_{2}(\vec{x}\,')^{*}U_{1}(\vec{x}\,')\right |^{2}- \sum_{\vec{x}\,'}^{} \left|U_{2}(\vec{x}\,')^{*}U_{1}(\vec{x}\,')\right |^{2}\right \}2\left | \alpha \right |^{2}\sinh^{2}(r) \nonumber \\
&&\qquad \qquad -\left \{ \left(\sum_{\vec{x}\,'}^{}U_{2}(\vec{x}\,')^{*}U_{1}(\vec{x}\,')\right )^{2}-\sum_{\vec{x}\,'}^{} \left(U_{2}(\vec{x}\,')^{*}U_{1}(\vec{x}\,')\right )^{2}\right \}\ \alpha^{\ast^{2}} \sinh(r) \cosh(r) + c.c.
\end{eqnarray}
\end{widetext}

For the special case of mode matching between squeezed light and local oscillator, camera is placed at the focal point of the lens and object is imaged directly on it. Hence object acts only as an intensity mask and does not add any phase to the transmitted light. Therefore, $T_1$ simplifies to a diagonal matrix comprised of 0s and 1s with the condition:
\begin{equation}
\hat{T}_{1}(\vec{x})= \left\{\begin{matrix}
1 & - & $no object$\\
0 & - & $object$.
\end{matrix}\right.
\label{eq:lsuO}
\end{equation}
\begin{figure}[t]
\centering
\includegraphics[width=8.5cm]{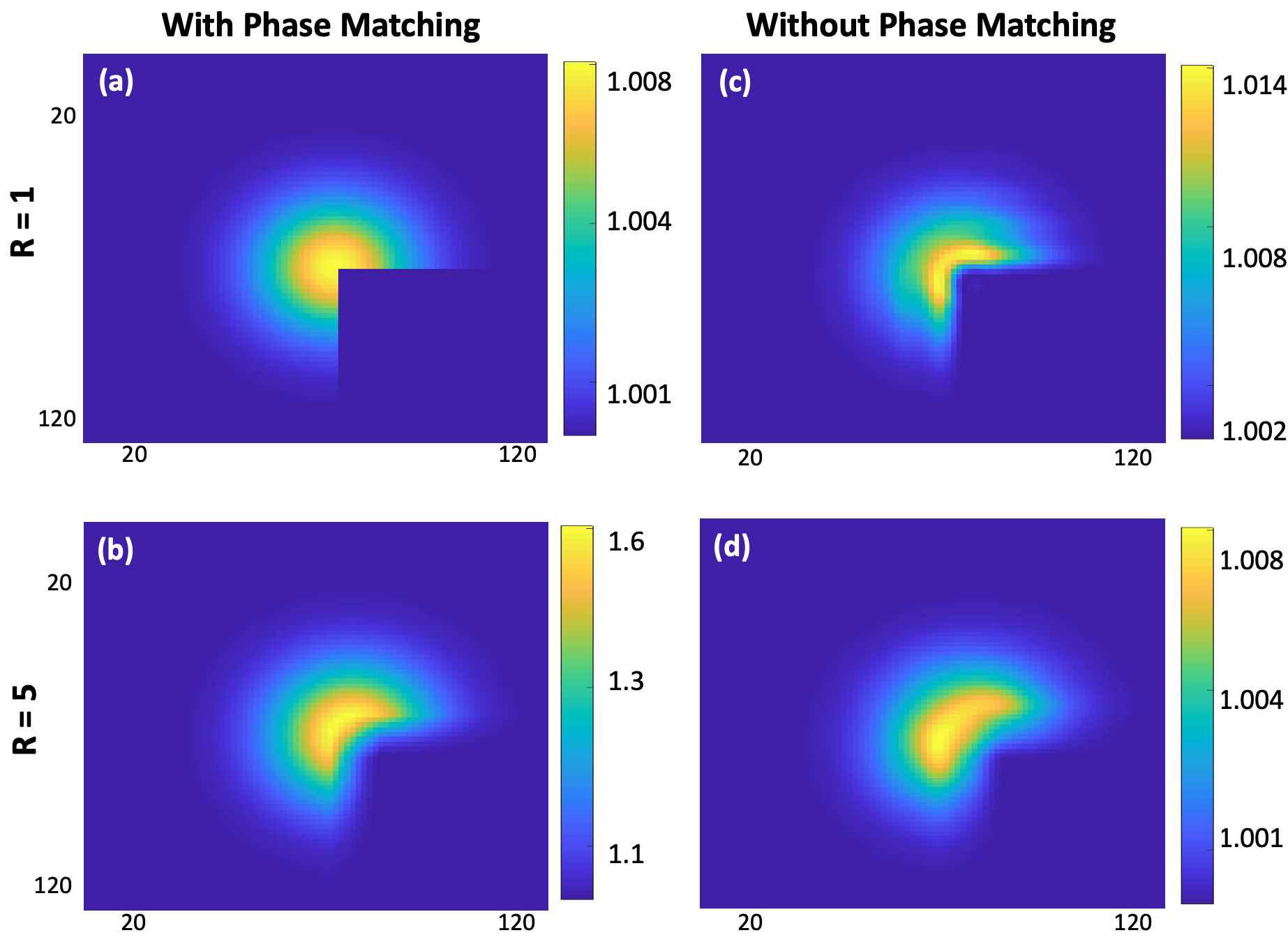}
\caption{
\label{fig:theory_noise_map}
The first column (a \& b) shows normalized noise maps of a mask in vacuum port calculated with phase matching condition. Binning of images (a) and (b) is done with detection areas of radius R =1 \& R=5 (in pixel units), respectively. The second column (c \& d) shows normalized noise maps of a mask in vacuum port calculated without phase matching condition. Binning of image (c) and (d) is done with detection areas of radius R =1 \& R=5 (in pixel units), respectively.}
\end{figure}
Furthermore, mode matching condition allows us to write $U_{1}(\vec{x})=T_{1}(\vec{x})\cdot U_{2}(\vec{x})$. This simplifies binned variance as
\begin{eqnarray}
&&\mathscr{V}^{(R)}_{th}(\vec{x})=|\alpha|^{2}\:\sum_{\vec{x}'}|U_{2}(\vec{x\,}')|^{2} + \\
&&|\alpha|^{2}\:(e^{2r}-1)\:\left(\sum_{\vec{x}\,'}T_1(\vec{x}\,')\left|U_{2}(\vec{x}\,')\right|^{2}\right)^{2},
\end{eqnarray}
and binned normalized variance as
\begin{equation}
V^{(R)}_{th}(\vec{x})=1+(e^{2r}-1)\frac{\left(\sum_{\vec{x}}T_1(\vec{x})\left|U_{2}(\vec{x})\right|^{2}\right)^{2}}{\sum_{\vec{x}}|U_{2}(\vec{x})|^{2}}.
\label{eq:lsuBVn}
\end{equation}
The validity of this approach is tested by calculating the noise map of a
map, similar to the second column in Fig.~3 in the main text.
128$\times$128 complex-valued map of field amplitudes, generated by
classical Fourier optics simulation of a Gaussian beam and a mask in the
path of a Gaussian beam is used for $U_{1}(\vec{x})$ and $U_{2}(\vec{x})$,
respectively. Figure~\ref{fig:theory_noise_map} shows the normalized noise maps calculated with and without phase-matching conditions for no binning (R=1) and binning (R=5) cases.
Normalized variance values greater than unity in the binned image, as opposed to shot noise limited variance obtained without any binning indicates the detection of anti-squeezed light demonstrated in experimental data. Our assumption of phase matching condition is supported by shot noise limited noise maps even after binning in Figure~\ref{fig:theory_noise_map}(d). The difference in R=1 images obtained by experiments and those by theory can be attributed to the various experimental sources of noise that were not considered in these calculations.

\end{document}